Mapping physiological suitability limits for malaria in Africa under climate change.

Running title: Climate change and malaria

Sadie J. Ryan[1,2,3,4]*, Amy McNally [6], Leah R. Johnson[7], Erin A. Mordecai[8], Tal Ben-Horin[9], Krijn Paaijmans[10], Kevin D. Lafferty [11,12]

*[1]Department of Geography, 3128 Turlington Hall, University of Florida, Gainesville, FL, 32611-7315, USA. sjryan@ufl.edu

[2] Emerging Pathogens Institute, 2055 Mowry Road, University of Florida, Gainesville, FL, 32610, USA

[3]Center for Global Health and Translational Science, Department of Immunology and Microbiology, Upstate Medical University, Syracuse, NY, 13210, USA

[4]School of Life Sciences, College of Agriculture, Engineering, and Science
University of KwaZulu-Natal, Durban, South Africa

[6]Hydrological Sciences Laboratory, NASA Goddard Space Flight Center, Greenbelt, MD 20771

[7]Division of Integrative Biology, University of South Florida, Tampa, FL 33620

[8]Department of Biology, Stanford University, Stanford, CA 94305

[9]Department of Marine and Coastal Sciences, Rutgers University, New Brunswick, NJ 08901

[10]Barcelona Centre for International Health Research (CRESIB, Hospital Clínic-Universitat de Barcelona), Barcelona, 08036, Spain

[11]Western Ecological Research Center, U.S. Geological Survey, Marine Science Institute, University of California, Santa Barbara, CA 93106

[12]Marine Science Institute, University of California, Santa Barbara, CA 93106

**Abstract**

We mapped current and future temperature suitability for malaria transmission in Africa using a published model that incorporates nonlinear physiological responses to temperature of the mosquito vector *Anopheles gambiae* and the malaria parasite *Plasmodium falciparum*. We found that a larger area of Africa currently experiences the ideal temperature for transmission than previously supposed. Under future climate projections, we predicted a modest increase in the overall area suitable for malaria transmission, but a net decrease in the most suitable area. Combined with human population density projections, our maps suggest that areas with temperatures suitable for year-round, highest risk transmission will shift from coastal West Africa to the Albertine Rift between Democratic Republic of Congo and Uganda, while areas with seasonal transmission suitability will shift toward sub-Saharan coastal areas. Mapping temperature suitability places important bounds on malaria transmissibility and, along with local level demographic, socioeconomic, and ecological factors, can indicate where resources may be best spent on malaria control.

**Keywords:** Malaria, climate change, physiological response, Africa

**Introduction**

Malaria causes an estimated 584,000 deaths a year, mostly due to *Plasmodium falciparum*, and mostly in sub-Saharan Africa (WHO 2014). This substantial health burden is anticipated to increase with changing climate (Shuman 2010; IPCC 2007a; Parham et al. 2015 and references therein). Many factors impact malaria burden, including climate, land use, socioeconomic conditions, and intervention efforts (Johnson et al. 2014). In particular, the temperature sensitivity of mosquitoes and parasites limits the transmission potential in a given location, and temperature has been shown to be an important predictor of malaria incidence in many areas (Pascual et al. 2006). Because insect and parasite physiology constrain malaria transmission to temperatures between 17-34°C (Mordecai et al. 2013), temperature places limits on the spatial and temporal distribution of malaria transmission. Climate induced shifts in distribution will impact the efficacy of intervention and vector control efforts (Siraj et al. 2014). As a result, accurately modeling how temperature limits transmission is essential to our understanding of the current and future distribution of malaria on the landscape (Snow, Marsh, and Le Sueur 1996).

Control measures can interrupt malaria transmission where climate conditions are suitable (Gething et al. 2010), making it difficult to predict climate suitability for transmission using statistical models of observed malaria transmission (human case data). An alternative is to use physiological (or process-based) models of transmission that explicitly link environmental drivers, in this case temperature, with transmission potential, to build maps of suitability. Mordecai *et al.* (2013) proposed a physiological model to assess the range of temperatures most suitable for malaria transmission. Specifically, the model incorporated the thermal ecology of anopheline mosquitoes (particularly *Anopheles gambiae*, the major vector of malaria in Africa) and *Plasmodium falciparum,* the major malaria parasite. This model included biologically-relevant unimodal physiological responses of all entomological and parasite life history traits to temperature, derived from empirical data. Including these explicit thermal responses reduced the predicted optimal temperature for malaria transmission by 6°C from previous model estimates – from 31°C to 25°C. Additionally, the predicted lower bounds for transmission decreased from around 20°C to 17°C, and the upper bounds decreased from 40°C to 34°C (Mordecai et al. 2013). Model predictions aligned closely with climate-matched field observations of EIR (the entomological inoculation rate, n=122), over a period of 40 years (Mordecai et al. 2013, Figure 2).

These differences in temperature limits should alter predicted spatial patterns of malaria transmissibility as the climate changes. More specifically, this model alters predictions of where climate both promotes and prevents transmission (Lafferty 2009; Rohr et al. 2011). Process-based models generate suitability predictions that are independent of observed prevalence data (rather than fitted to it), and not confounded by control efforts or socio-economic factors. Thus, this model provides a baseline prediction for the impacts of temperature in determining large-scale patterns of malaria transmissibility, to which more local effects (e.g. water availability, topography, or control measures) can be added to improve predictions.

Based on the Mordecai *et al.* (2013) model we mapped malaria temperature suitability, seasonality, and transmission duration, under current and predicted future climate scenarios, incorporating restrictions on moisture availability. We then combined these with human population density estimates to emphasize changes in the relative risk of changing temperatures to human populations living in areas suitable for malaria transmission. We did not measure absolute transmission risk because other factors, including socioeconomic conditions, urbanization, and control measures affect the realized transmission rate (Gething et al. 2010). We mapped the relative intensity of conditions for transmission because when conditions are highly suitable for transmission, transmission rate may not strongly correlate with prevalence (Beier, Killeen, and Githure 1999), but will likely be mediated by human exposure behavior.

**Materials and Methods**

*Mapping thermal suitability*

The Mordecai et al. (2013) model measured transmission using $R_0$, the number of secondary infections expected when an infected individual enters a fully susceptible population. The model is an extension of the classic Kermack-McKendrick model, and $R_0$ is given by:

$$R_0 = \sqrt{\frac{M}{Nr}\frac{a^2 bc\, exp(-\mu/PDR)}{\mu}}$$

with $N$ the human population size, a is the per-mosquito biting rate, or mean oviposition time$^{-1}$, $bc$ the vector competence, $\mu$ the mosquito mortality rate, PDR the parasite development rate, and the mosquito population size $M$ (which is itself determined by mosquito fecundity, development rate, and adult and egg/juvenile survival). All vector and parasite traits are assumed to depend on temperature and are modeled with hump-shaped responses fit to laboratory data. Human population density and human recovery rate ($r$) from malaria are assumed to be temperature independent. When these are combined in the model, the result is a prediction curve of how $R_0$ should vary across temperatures (Mordecai et al., 2013).

We describe the predicted proportion of optimal $R_0$ using quantiles capturing Most (top 25%), High (25-50%), Moderate (50-75%) and Marginal (75-100%) suitability. These quantiles describe the proximity of temperatures to the optimal transmission temperature, i.e., the relative transmission risk, which we refer to as relative $R_0$. The top three quantiles (0-75%) correspond to temperatures that promote malaria, whereas the last (marginal) likely limits transmission. To demonstrate the effects of nonlinear thermal responses, we compared the results with models that used the thermal responses from a previous mechanistic transmission model (Parham and Michael 2010).

The thermal response models were used to predict relative $R_0$ as a function of temperature and mapped onto temperature at 0.01°C increments using a geographic WGS84 (decimal degree) projection map of the African continent. Because moisture availability also limits mosquito survival, we masked areas that are too dry for too long for mosquitoes to develop and survive, based on the Normalized Difference Vegetation Index (NDVI), sensu Suzuki *et al.* (2006), (see below). We then queried the map for suitability quantiles and calculated their areas. We used program R (R Core Team 2014), for our analyses; specifically, we used the packages raster, sp, maptools and rgdal, to import and manipulate raster data as R objects, and rasterVis, lattice, latticeExtra, and ColorSpace to create our visualizations.

*Climate data*

We used global monthly mean temperatures for the current climatic period (to 2000) calculated at a 30 arc-second resolution from WorldClim (Hijmans et al. 2005), and clipped to the African continent. For future climate predictions, we used HadCM3 climate projections for 2020, 2050 and 2080, and selected SRES A1B emission scenarios, downscaled using the delta method (Ramirez and Jarvis 2008). The A1B scenario falls in the center of projections of anthropogenic emissions, and makes a general assumption of continuing globalization (IPCC 2007b), while avoiding extremes. The delta method uses the long-term monthly mean of temperature generated by the HadCM3 global climate model to calculate anomalies for the generated time series (Ramirez and Jarvis 2008), which are applied to higher resolution temperature surfaces from WorldClim. We chose this method for its simplicity and sufficiency for continental level inter-model comparison. We aggregated the 5 arcmin (~0.083°), data to 0.1° to match the data used to project the $R_0$ models.

*Aridity mask*

We defined moisture availability using thresholds of vegetation greenness, similar to the approach used by Guerra *et al.* (2008), and following work by Suzuki *et al.* (2006). Recent work by Baeza *et al.* (2011) has shown that NDVI is a good predictor of malaria prevalence, typically better than rainfall when irrigation levels are low to moderate. Using NDVI allowed us to retain pockets of suitable transmission conditions such as river tributaries or irrigated land where transmission may be supported, while eliminating areas that are too dry for too long to support mosquito populations, regardless of temperature regime (Gething et al. 2011; Guerra et al. 2008; Guerra et al. 2010). Differences in rainfall within sufficiently humid regions may impact transmission, but NDVI has been shown to be a stronger predictor than precipitation for large geographic areas (Baeza et al. 2011). We created pixel-wise monthly mean NDVI values from the USGS eMODIS product (Jenkerson, Maiersperger, and Schmidt 2010), using dekadal (10-day interval) data from 2001-2011 for the African continent from the FEWS NET (Famine Early Warning System Network) data portal (http://earlywarning.usgs.gov/fews/africa/index.php), aggregated from 250m to 0.1 degree resolution. We masked out pixels that did not have at least two consecutive months of NDVI above 0.125; this is slightly more conservative than the 0.1 limit of Suzuki *et al.* (2006).

*Population risk projections*

To estimate the population at risk for transmission, we used population projection data for 2015 from the Gridded Populations of the World (GPW) ver 3.0 (2005), and the Global Rural Urban

Mapping Project (GRUMP) Urban extents layer (2011). We resampled the projected malaria model and urban extents to the GPW resolution (2.5 arc-minutes) using bilinear interpolation. To visualize the impact of malaria shift and population density, we first scaled malaria season duration (0-12 months) to a 0-1 scale, and log transformed population density, for ease of visualization. Malaria transmission ceases to be a function of the environment in high density, urban areas (Gething et al. 2011), so we excluded these from our analysis with the GRUMP data, by simply masking out the areas described as urban extents.

## Results

### *Current suitability*

We illustrate the seasonal changes in malaria transmission suitability as maps of relative $R_0$ for each month of the year (Figure 1). We contrast this with an illustration of the same maps generated for **a** previous model (Parham and Michael, 2010; Supplemental Figure 1). We compared the land area in each suitability category each month between the two models (Figure 1 inset, Supplemental Figure 1 inset). The area currently containing the most suitable transmission (top 25%) by month in the map of Mordecai et al. (2013) ranges from 5.8 to 10.9 million km$^2$ (Figure 1 inset). In the previous model, this ranged from 13,000 km$^2$ to 4.9 million km$^2$ (Supplementary Figure 1); this difference is due to the large expanse of moderate suitable temperatures captured in the new model.

### *Duration of the transmission season*

Much of Africa is predicted to currently be at least marginally suitable for transmission year-round (Figure 2b), while much of Central Africa is in the Most suitable quantile year-round (Figure 2a). This closely matches independent predictions by MARA (Mapping Malaria Risk in

Africa) (Tanser, Sharp, and le Sueur 2003) and MAP (Malaria Atlas Project) (Gething et al. 2010; Gething et al. 2011).

*Predicted future suitability*

The area projected to be suitable for any level of malaria transmission for one or more months modestly increased over time in the Mordecai *et al.* (2013) model (Figure 3). This differs considerably from predictions from the prior model, which predicts steady increase (Fig S2). In contrast, current highly suitable areas such as central and western Africa (Figure 4) are predicted to become less suitable with warming.

In the future scenarios, the area with temperatures most suitable for year-round (12 months) transmission contracted (Figure 2a vs. 2c). In particular, there was a reduction in the total land area with temperatures suitable for high year-round transmission. The hotspot of most suitable temperatures year-round shifted from central Africa to the Albertine Rift region in the East, and to Angola, Gabon and Cameroon in the West by 2080. Western Africa, particularly along the coast of Ghana, ceased to have optimal temperatures for year-round transmission. Future temperatures are predicted to promote a mix of episodic Most and High transmission due to warmer temperatures exceeding the upper limit of 34ºC for transmission. In contrast, high seasonal transmission potential (4-8 months) expanded throughout Southern Africa and Madagascar (Figure 2).

*Population risk mapping*

Our population risk map highlights hotspots where high population density overlaps with long season length with most transmission potential (Figure 4). Specifically, the current year-round most suitable and densely populated Ghana coast of western Africa was predicted to become more seasonal and less suitable for transmission, while this high-density, year-round most suitable hotspot is predicted to shift to the Albertine Rift border of Western Uganda (Figure 4).

**Discussion**

Maps based on a model with realistic assumptions about thermal physiology of anopheline mosquitoes made novel predictions about the relationship between climate change and malaria transmission suitability. Specifically, temperatures across a large area of Africa are currently more suitable for malaria transmission than previously estimated. Because this revised optimal temperature (25°C) is much closer to current temperature conditions across much of Africa, future warming may reduce transmission suitability in regions where transmission is currently high. Still, temperatures across large parts of the continent will become more suitable for transmission overall under future climate scenarios, and the areas where low temperature completely prevents malaria transmission will contract. These contrasting patterns arise from the unimodal relationship between malaria transmission and temperature.

Our map shows that most of Africa now has temperatures suitable for malaria transmission to some degree, throughout the year. In fact, some regions are already beyond the thermal optimum for transmission, particularly just south of the Sahara (Figure 1, Figure 2b). Suitability also varies seasonally (Figure 1): in March through June, the total most suitable area (top quantile) is smaller (Figure 1, inset) because it is too hot in the north (Figure 1, red color in maps) and too cold in the south (Figure 1, blue color in maps). By contrast, much of the continent is most suitable from November through February. Further, the new maps capture the current observed transmission in cooler regions where malaria is an ongoing concern, such as the KwaZulu-Natal region of South Africa (Kleinschmidt et al. 2001; Barnes et al. 2005). A map based on previous linear thermal responses predicted that most regions of Africa are currently cooler than optimum for transmission, with transmissibility peaking in April (Supplementary Figure 1).

By incorporating empirical findings that malaria transmission is optimal under moderate temperatures, and transmission potential declines at hot temperatures, our model generates nuanced predictions of current and future temperature suitability. For example, although the model predicts a small increase in the amount of geographic area with temperature suitable for any transmission by 2080, it also predicts a reduction in the area with temperature most suitable for transmission. Areas predicted to become newly suitable for transmission could face challenges to vector control and health infrastructure to mitigate increased transmission risk. Moreover, some of these predicted areas also coincide with centers of high population density, such as Western Uganda, suggesting large predicted geographic shifts in the number of people living where conditions are most suitable for transmission Africa.

Our seasonal duration model of the most suitable transmission quantile was visually similar to the MARA map (http://www.mara.org.za/pdfmaps/AfMonthsRisk.PDF, *accessed May 15, 2013*). In contrast, our 100% range map of suitability predicts a much larger area of 12-month suitability, because it captures all areas where malaria transmission is not excluded for a full 12 months. Our map predictions differ from the MARA map in several regions, specifically in the Horn of Africa and in Southern Africa; we find suitability where it is not predicted by MARA. These differences almost certainly reflect factors other than temperature, including different assumptions about humidity. In the MARA model specifications, full suitability in southern Africa (i.e., south of 8° latitude) was only achieved if an area experienced temperatures above 6°C and 80 mm or more of rain for at least *five* months (Guerra 2007) – a strict multivariate limit. This limit is used to focus MARA maps on endemic malaria areas, which is only a subset of what we considered. Other recent maps that focus specifically on areas in Africa suitable for epidemic (episodic) malaria (Grover-Kopec et al. 2005) match well with our predicted areas of lower transmission (lower suitability, short season). Our current predictions (Fig. 2a) correspond broadly with current *Plasmodium falciparum* parasite prevalence measured in children ages 2-10 (see (Noor et al. 2014), Figure 2). A temperature-only suitability map based only on the duration of sporogony for *P. falciparum*, developed by Guerra (2007) as a step in a larger modeling process, indicates that all of Africa could potentially be suitable for the malaria parasite (see (Guerra 2007) Figure 6.6), in line with our 'marginally suitable' or largest range limit. Our model also broadly corroborates predictions of current and future *Anopheles gambiae* range in Africa using an ecological niche model (Peterson 2009). As each malaria or vector mapping project discussed used slightly different criteria, we cannot expect these maps to correspond exactly. Other approaches, such as those that rely on empirical transmission data, could inadequately capture conditions that influence transmission in areas with human populations that are too small to sustain malaria, or where data collection effort is too low to capture intermittent transmission. The challenges in understanding why each approach results in different predicted patterns highlight the need to make the mapping process more transparent across studies so they can be more directly compared (Johnson et al. 2014).

Our maps suggest that projected climate change could lead to a large decrease in the population at highest risk for malaria transmission. Future shifts in population density will affect the distribution of people living where temperatures are suitable for malaria transmission, and this distribution might also respond directly to changes in climate or malaria transmission. However, we emphasize that the Mordecai *et al.* (2013) model predicts that optimum transmission temperatures are cooler than in previous models. Given the predicted geographic shifts of these moderate temperatures, it is likely that the optimal transmission temperature and high population density will continue to coincide. Finally, our predicted transmission suitability under climate change is limited to the predictive quality of the climate model we use; understanding the importance of assumptions underlying climate change scenarios would require repeating the analyses described here under different climate regimes.

We still need a better understanding of the behavior and physiology of local mosquito and parasite populations and their climate space. One thermal performance curve cannot capture the full breadth of thermal response behavior of all mosquito and parasite populations across Africa. Ectothermic populations differ in their responses to temperature within species, due to adaptation of local populations to prevailing local conditions (Angilletta Jr, Niewiarowski, and Navas 2002), which may affect their capability to cope with climatic change (Sternberg and Thomas 2014). Moreover, the actual temperatures that vectors and parasites experience can differ considerably from standard weather-station data (Cator et al. 2013), in part because mosquitoes can behaviorally thermoregulate (Blanford, Read, and Thomas 2009) and avoid extreme temperatures (Kirby and Lindsay 2004). Temperature variation (e.g., daily temperature variation) may also be important for making more accurate local predictions, but we do not expect it to affect our general predictions for relative suitability. While daily fluctuations have been shown to affect mosquito and parasite development (Paaijmans et al. 2010), this occurs mostly at the tails (upper and lower boundaries) of the range. Blanford et al. (2013) found that these temperature variations averaged over a month performed just as well as daily fluctuations for estimating biological rates. Nonetheless, local studies on mosquito physiology and behavior, combined with microclimatic measurements, will allow us to determine how climate, climate variability and climate extremes eventually affect disease transmission at a finer scale (Altizer et al. 2013).

Our approach focuses on determining how temperature impacts broad geographic patterns of malaria transmission potential. Spatial regression models incorporating many other environmental and socio-economic covariates such as atmospheric humidity, rainfall, or bed-net availability could further enhance predictions at a local scale. Our model assesses the direct influence of temperature on transmission mechanistically. Thus, our results complement statistical models derived from field data to inform why malaria is excluded in some regions, and highly prevalent in others. It can also serve as a baseline for predicting future suitability, as mechanistic models are useful for predicting outside the range of current data than traditional regression based approaches (Bayarri et al. 2009; Rogers and Randolph 2006). This framework allows us to disentangle the effects of other important factors, such as land use and disease control efforts (Gething et al. 2010), to better predict transmission risk in the field. We note that the timescale at which climate change may affect the distribution of suitability does not necessarily match that of health intervention planning. However, this suitability map can inform current allocation of intervention under current climate conditions by highlighting areas that perhaps were not previously identified as seasonally suitable, or by distinguishing endemic suitability and epidemic suitability. Given that the types of intervention and control differ in these contrasting scenarios, and that resistance can emerge in more episodic areas, we think this map is of value to planning. Most importantly, with accurate suitability predictions and appropriate planning for potential geographic shifts in risk, an increase in transmission suitability does not have to increase malaria cases.

**Acknowledgments**

This work was conducted as a part of the Malaria and Climate Change Working Group supported by the Luce Environmental Science to Solutions Fellowship and the National Center for Ecological Analysis and Synthesis, a Center funded by NSF (Grant #EF-0553768), the University of California, Santa Barbara and the State of California. Greg Husak and Bobby Gramacy provided input on data acquisition. EAM was supported by an NSF Postdoctoral Research Fellowship in Biology (DBI-1202892). Any use of trade, product, website, or firm names in this publication is for descriptive purposes only and does not imply endorsement by the U.S. Government.

**Author Disclosure Statement**

No competing financial interests exist

**Figures**

Figure 1. Areas with mean monthly temperatures suitable for malaria transmission as predicted by the model of Mordecai et al. (2013). Suitability is shown under the current climate (see Methods), on a blue-red scale from too cold for transmission (blue) through peak optimal transmission (white), to too hot for transmission (red). The aridity mask, where the area is unsuitable for mosquito development is shown in grey, centered on the Sahara desert. The inset panel shows the area of land (in km$^2$ x10$^6$) in each month, within each quantile of transmission suitability, as described in the text.

Figure 2. Duration of temperatures suitable for malaria transmission. Duration of malaria season is given as the number of continuous months suitable for transmission, for the top 25% of suitability (most – panels a and b) and the full 100% range (suitable - panels b and d), as predicted by the Mordecai et al. (2013) model, in current (a and b), and 2080 (c and d) climate conditions.

Figure 3. The area of land on the African continent contained within each quantile for temperature suitability for transmission. Suitability area quantiles averaged across months are shown as predicted by the Mordecai et al. (2013) model for current and future (A1B) climate scenarios: 2020, 2050, 2080, as described in Methods and Materials.

Figure 4. Potential population at risk for most suitable malaria transmission temperatures (top 25%), under a. Current, b. 2020, c. 2050, d. 2080 climate projections, where risk is measured as a combined scale of change in transmission suitability duration (0-12 months), scaled 0-1, and population density (log-transformed). Areas circumscribed in red describe 'hotspots' of coincident high population density and year-round most optimal transmission.

Figure S1. Monthly temperature suitability for malaria transmission, as predicted by the thermal responses in Parham and Michael (2010). Temperature suitability is shown under the current climate (see methods), on a blue-red scale from too cold for transmission (blue) through peak optimal transmission (white), to too hot for transmission (red). The aridity mask, where the area is unsuitable for mosquito development is shown in grey, centered around the Sahara desert. The inset panel shows the area of land (in $km^2$ $x10^6$) in each month, within each quantile of transmission suitability, as described in the text.

Figure S2. The area of land on the African continent with temperatures within each transmission suitability quantile based on the thermal responses in Parham and Michael (2010). Transmissibility area quantiles averaged across months are shown as predicted by the thermal responses for current and future (SRES A1B) climate scenarios: 2020, 2050, 2080, as described in the Methods.

**Figure 1.**

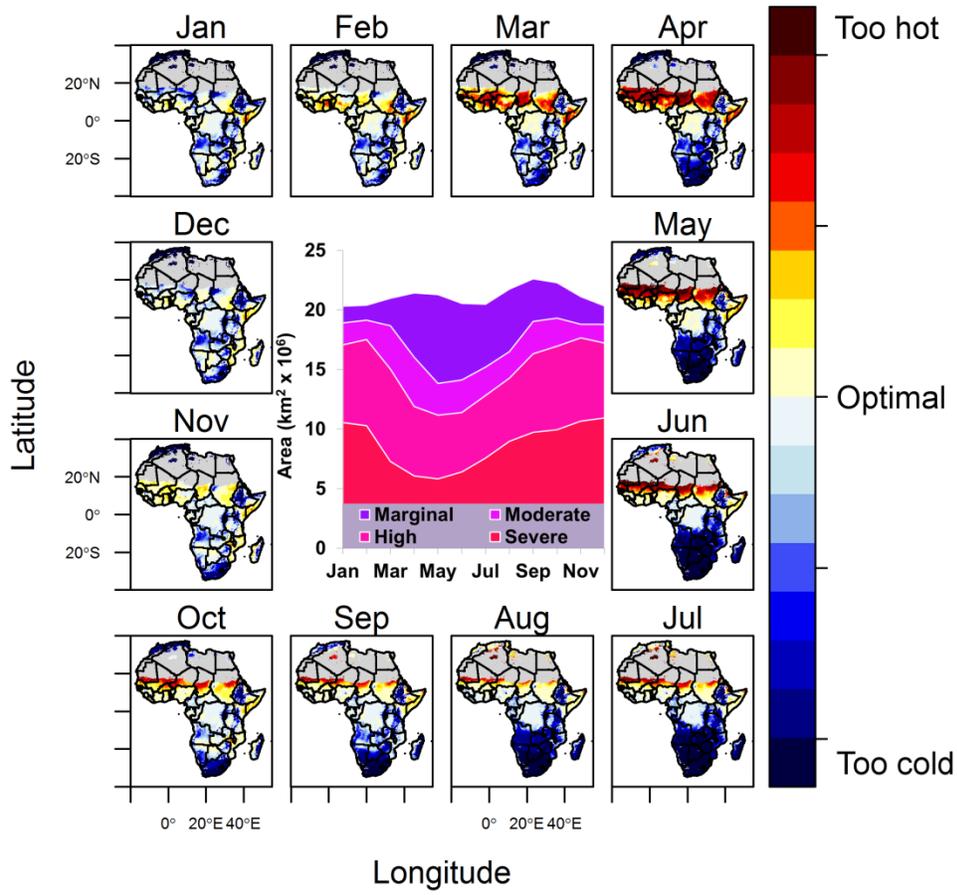

**Figure 2**

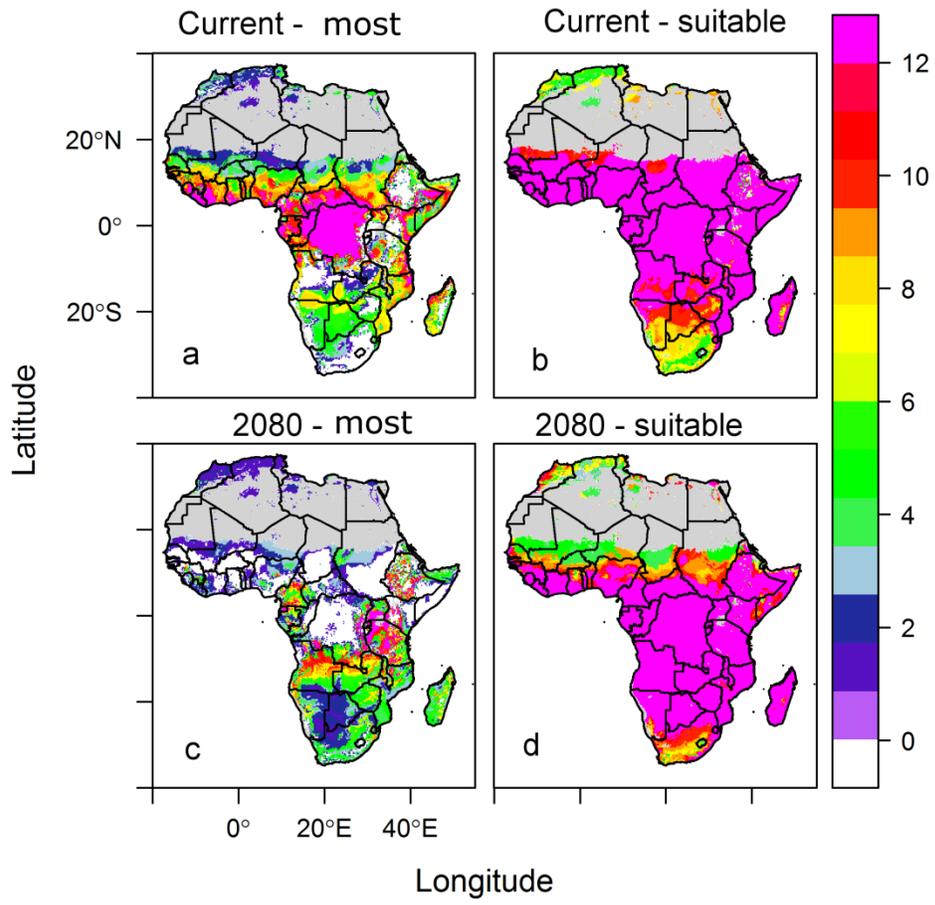

**Figure 3**

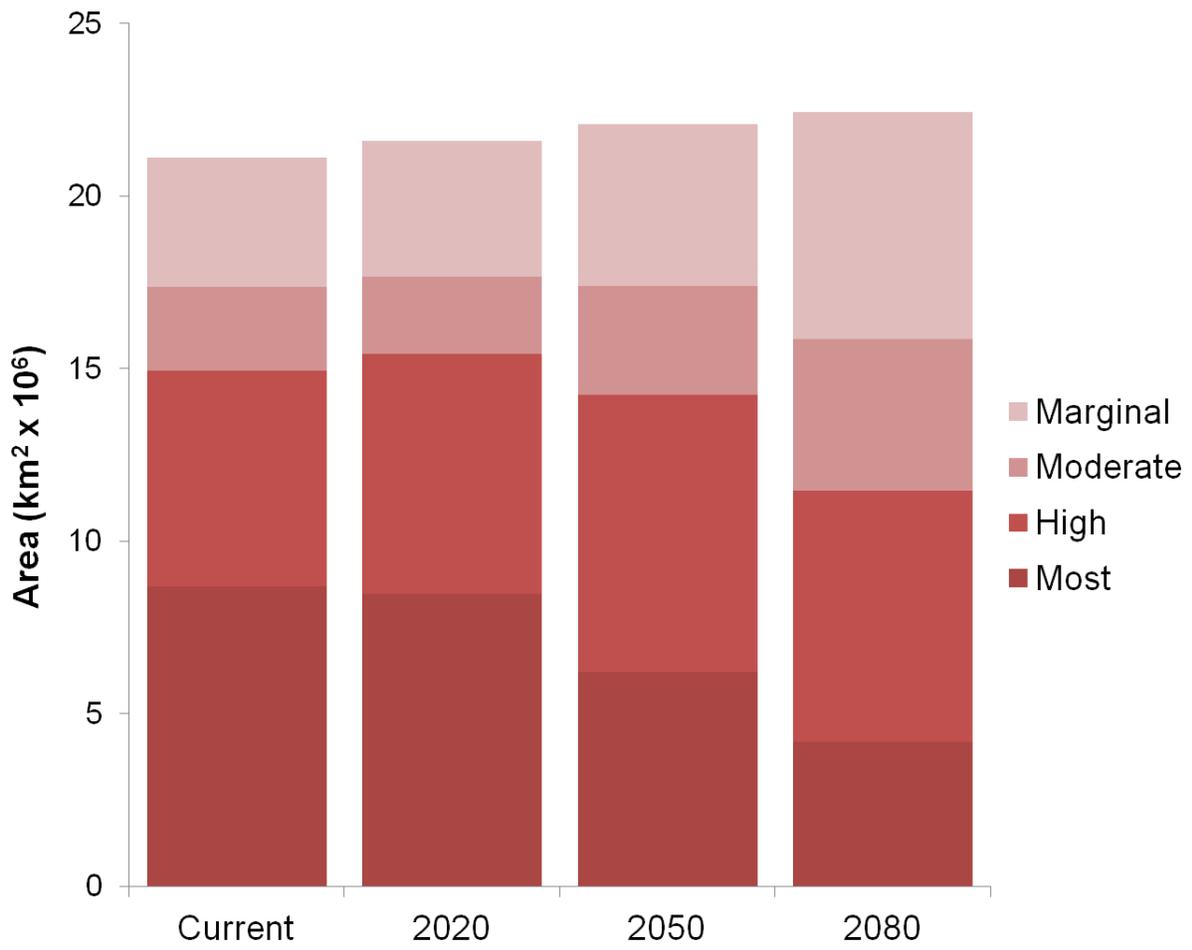

**Figure 4**

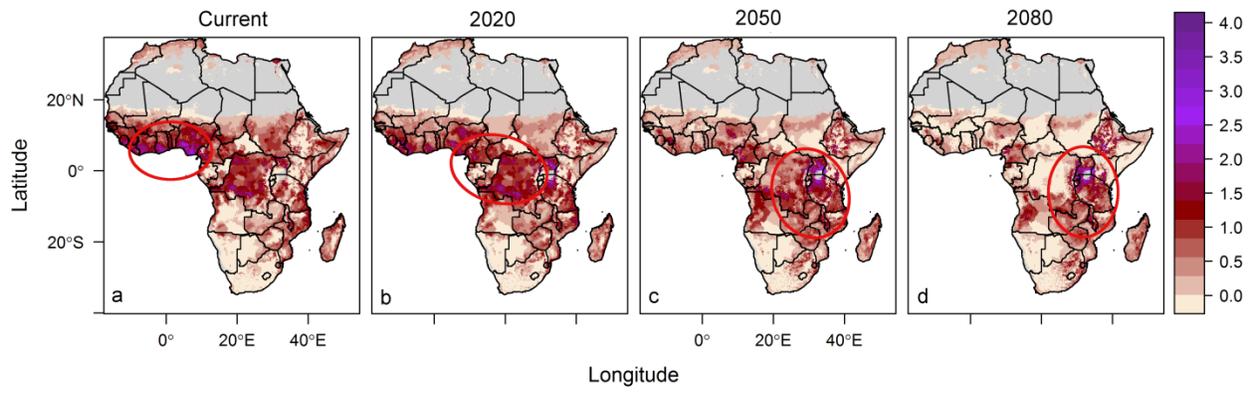

**Figure S1**

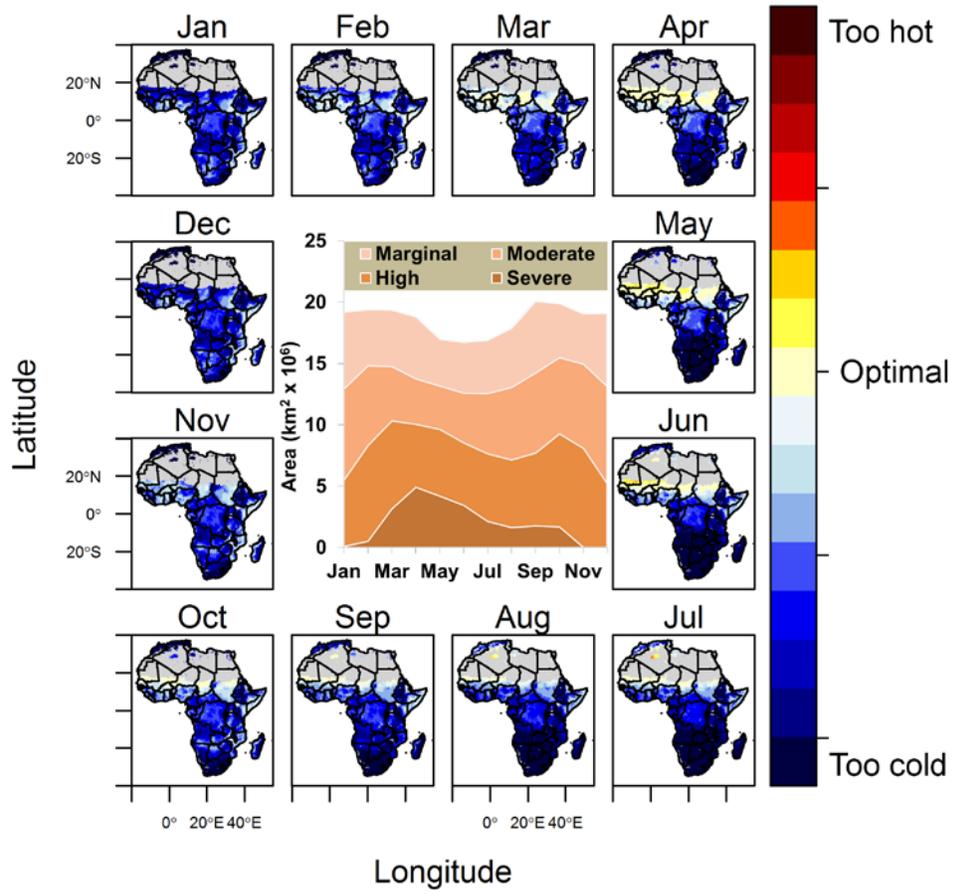

**Figure S2**

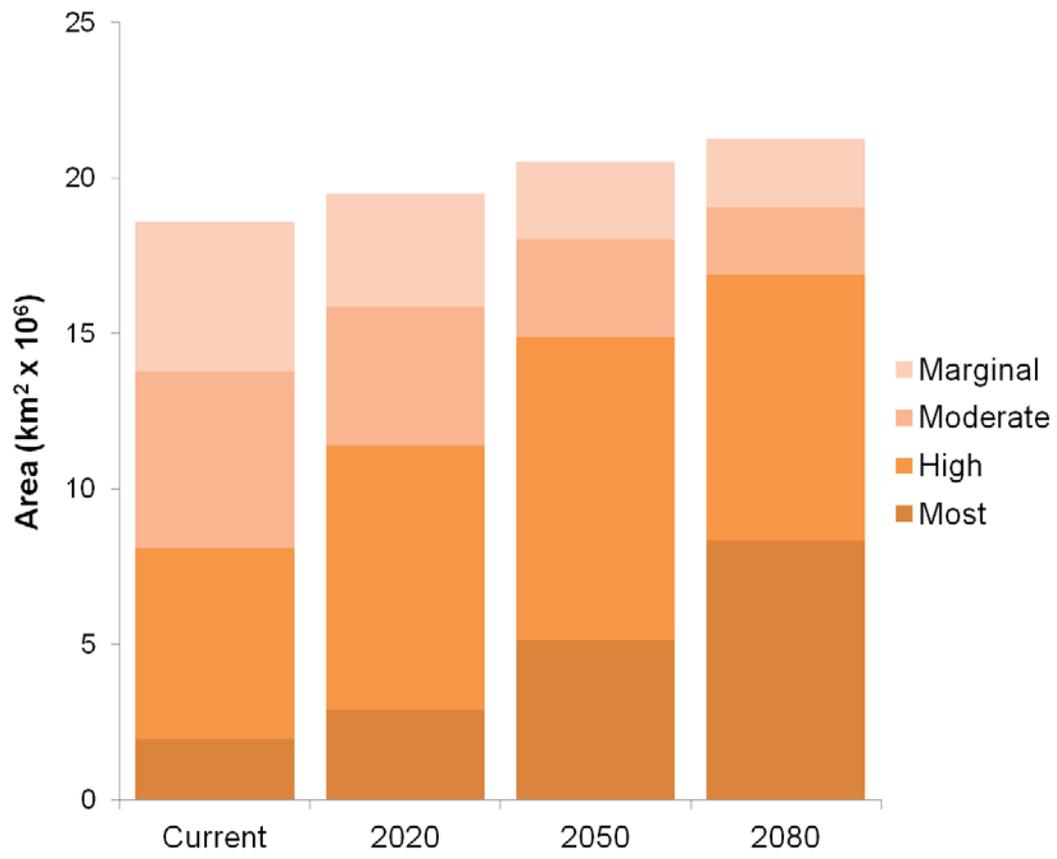